%% file: main.tex
\pdfoutput=1
\documentclass[aps,prl,reprint,nofootinbib,superscriptaddress,longbibliography]{revtex4-2}
\input{macros}
\hypersetup{pdftitle={Vacuum correlation measures as CFT partition functions}}

\begin{document}

\title{Vacuum correlation measures as CFT partition functions}

\author{Hewei Frederic Jia}
\email{heweifred@gmail.com}
\affiliation{Institute for Advanced Study, Tsinghua University, Beijing, 100084, China}

\begin{abstract}
We show that in any $d$-dimensional CFT, the vacuum sandwiched R\'enyi mutual information (SRMI) of integer order $n$ between two disjoint balls maps to a CFT partition function on a manifold of topology $\mathop{\#}^{n-1}(S^1\times S^{d-1})$, with moduli fixed by the ball configuration and single-ball modular flow. At $n=2$, it is given by the thermal partition function on $S^{d-1}$, upper-bounding the von Neumann mutual information. For $n>2$, SRMI diverges at finite separation due to a hot-spot limit of the partition function. Explicit CFT evaluations agree well with density matrix computations.
\end{abstract}

\maketitle

\emph{Introduction}---Quantifying vacuum correlations between disjoint regions $A$ and $B$ in a conformal field theory (\CFT{d}) is a central problem across quantum many-body physics, quantum field theory, and holography. The canonical measure is the von Neumann mutual information (MI) $I(A:B) = D(\rho_{AB}\Vert \rho_A\otimes\rho_B) = S(A) + S(B) - S(AB)$. Because it is a relative entropy $D(\rho\Vert\sigma) = \Tr[\rho(\log\rho - \log\sigma)]$, it is nonnegative and obeys data processing, yet remains difficult to evaluate at finite separation in interacting theories even in two dimensions. Although R\'enyi entropies $S_n$ are more accessible via replica constructions, the naive combination $S_n(A)+S_n(B)-S_n(AB)$ fails positivity and data processing. A proper R\'enyi mutual information should instead replace relative entropy by a R\'enyi relative entropy. A distinguished choice is the sandwiched R\'enyi relative entropy~\cite{MullerLennert,WildeWinterYang,FrankLieb,Lashkari}, which at R\'enyi order $\alpha$ is defined by
\begin{equation}
  \SRel{\alpha}(\rho\Vert\sigma)
  = \frac{1}{\alpha-1}\log\Tr\!\left[\left(\sigma^{\frac{1-\alpha}{2\alpha}}\rho\,\sigma^{\frac{1-\alpha}{2\alpha}}\right)^\alpha\right].
  \label{eq:sandwiched-def}
\end{equation}
Unlike the Petz--R\'enyi divergence, which violates data processing above $\alpha = 2$, $\SRel{\alpha}$ obeys data processing for all $\alpha \ge 1/2$ and increases monotonically with $\alpha$. Replacing $D$ by $\SRel{\alpha}$ defines the sandwiched R\'enyi mutual information (SRMI)~\cite{Scalet,KNV},
\begin{equation}
  \SRMI{\alpha}(A:B) = \SRel{\alpha}(\rho_{AB}\Vert \rho_A\otimes\rho_B).
  \label{eq:srmi-def}
\end{equation}
It reduces to MI as $\alpha \to 1$. In this Letter we focus on integer orders $\alpha = n \ge 2$.

In field theory, the negative fractional powers $\sigma^{\frac{1-n}{2n}}$ obstruct standard replica constructions, leaving SRMI for disjoint regions at finite separation inaccessible thus far. Here we take a different route using the radial sewing expansion of $\rho_{AB}$ between two disjoint balls, introduced in Ref.~\cite{CompanionMI} for the long-distance expansion of MI. Each term in this expansion contains half-powers $\rho_A^{1/2}$ and $\rho_B^{1/2}$ generated by single-ball modular flow. Inside $\SRMI{n}$, they absorb the negative powers $\sigma^{\frac{1-n}{2n}}$, leaving positive powers $\rho_{A,B}^{\frac{1}{2n}}$ that implement Euclidean modular rotations and space $n$ insertions evenly around each of two copies of compactified Euclidean spacetime $S^d$. Contracting the radial state sums across $n$ cylinders then glues the two $n$-punctured copies of $S^d$ into a smooth manifold $M_{n-1}(\xi) \simeq \mathop{\#}^{n-1}(S^1\times S^{d-1})$. For integer R\'enyi order $n \ge 2$, SRMI thus equals the vacuum-normalized CFT partition function on $M_{n-1}(\xi)$.

At $n = 2$, SRMI reduces to the thermal partition function on $S^{d-1}$, with an effective temperature depending on the ball configuration. This yields a thermal upper bound on MI that strictly improves a similar bound in Ref.~\cite{CeyhanFaulknerThermal}. In holographic CFTs, this implies that the Hawking--Page transition of second SRMI precedes the Ryu--Takayanagi phase transition of MI. For $n \ge 3$, SRMI diverges at a finite separation, with the singular behavior generically controlled by a hot-spot limit of the partition function on $M_{n-1}(\xi)$~\cite{BLOSD}. We also find excellent agreement between CFT partition function evaluations and density matrix computations in explicit examples.

\emph{Sewing construction and CFT partition function}---By conformal invariance, two disjoint equal-time balls in $\mathbb{R}^d$ can be mapped to concentric spheres of radii $\xi$ and $1$:
\begin{equation}
  A = \{(0,\bm{x}) : |\bm{x}| < \xi\}, \qquad
  B = \{(0,\bm{x}) : |\bm{x}| > 1\},
  \label{eq:regions}
\end{equation}
with $0 < \xi < 1$ and four-point cross ratio
\begin{equation}
  \eta = \frac{4\xi}{(1+\xi)^2}.
  \label{eq:cross-ratio}
\end{equation}

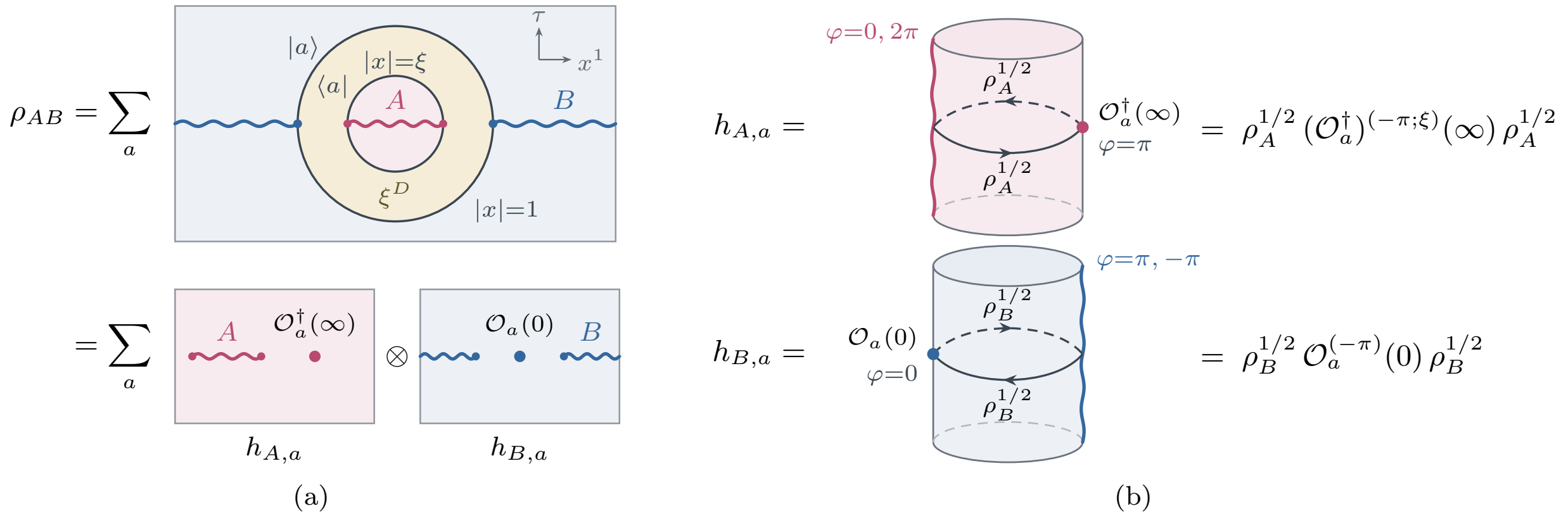
\begin{figure*}[tbp]
  \centering
  \input{figures/radial-preparation}
  \caption{Radial sewing expansion of $\rho_{AB}$. (a) Splitting the Euclidean path integral on $|x|=\xi$ and $|x|=1$ isolates the annulus $\xi^D$ (gold), expressing $\rho_{AB}$ as a sum of single-cut path integrals $h_{A,a}$ and $h_{B,a}$ [Eq.~\eqref{eq:radial-factorization}]. (b) On the modular cylinder $S^1_\varphi\times\mathbb{H}^{d-1}$, half-period modular flow pulls each insertion onto its cut, giving the sandwiched form [Eq.~\eqref{eq:rhoAB-exp}].}
  \label{fig:radial-prep}
\end{figure*}

We start by reviewing the radial sewing expansion of density matrix $\rho_{AB}$~\cite{CompanionMI}. The joint density matrix is prepared by a Euclidean path integral on $\mathbb{R}^d$ with cuts at $A,B$. We split the Euclidean path integral into three regions separated by the concentric spheres $|x| = \xi$ and $|x| = 1$. The annulus region $\xi < |x| < 1$ carries the radial propagator $\xi^D = \sum_a \xi^{\Delta_a}\ket{a}\bra{a}$ in an orthonormal dilatation eigenbasis. The remaining two parts of the path integral are contracted with $\bra{a}$ at radius $|x|=\xi$, and with $\ket{a}$ at radius $|x|=1$, respectively. By state-operator correspondence, the state $\ket{a}$ at radius $|x|=1$ is prepared by inserting $\cO_a(0)$ at the origin, and the state $\bra{a}$ at radius $|x|=\xi$ is prepared by inserting the radial conjugate operator $\xi^{-\Delta_a}\cO_a^\dagger(\infty)$ at infinity, which cancels the annulus weight $\xi^{\Delta_a}$. This writes $\rho_{AB}$ as sum of single-cut path integrals [Fig.~\ref{fig:radial-prep}(a)]:
\begin{equation}
  \rho_{AB} = \sum_a h_{A,a} \otimes h_{B,a},
  \label{eq:radial-factorization}
\end{equation}
with single-region operators $h_{A,a} = \Tr_{A^c}[\ket{0}\bra{0}\,\cO_a^\dagger(\infty)]$ and $h_{B,a} = \Tr_{B^c}[\cO_a(0)\,\ket{0}\bra{0}]$.

To express $h_{A,a}$ and $h_{B,a}$ as operators sandwiched by $\rho_{A,B}^{1/2}$, we use the geometric modular flow of a single ball $\mathcal{B}_R = \{(0,\bm{x}) : |\bm{x}| < R\}$~\cite{CHM}. Conformally mapping $\mathbb{R}^d \setminus \partial\mathcal{B}_R$ to the modular cylinder $S^1_\varphi \times \mathbb{H}^{d-1}$ places $\mathcal{B}_R$ at $\varphi = 0$ and its complement $\mathcal{B}_R^c$ at $\varphi = \pi$, with modular rotation $\varphi \mapsto \varphi + \theta$ implemented by
\begin{equation}
  \begin{aligned}
    U_\theta^{(R)}
    &= \exp\!\left[\frac{\theta}{2}\!\left(R P_0 + R^{-1} K_0\right)\right]
     = R^D U_\theta R^{-D} \\
    &= \rho_{\mathcal{B}_R}^{-\theta/(2\pi)} \otimes \rho_{\mathcal{B}_R^c}^{\theta/(2\pi)},
  \end{aligned}
  \label{eq:Utheta}
\end{equation}
where $U_\theta \equiv U_\theta^{(1)}$ and its coordinate action on $x = (\tau,\bm{x})$ is
\begin{equation}
  U_\theta^{(R)} \cdot x
  = \frac{x + \bigl[(\cos\theta - 1)\tau + \frac{1}{2R}\sin\theta\,(R^2 - x^2)\bigr] e_0}
         {\cos^2(\theta/2) - \frac{\tau}{R}\sin\theta + \frac{x^2}{R^2}\sin^2(\theta/2)},
  \label{eq:gtheta}
\end{equation}
with $e_0 = (1,\bm{0})$. We denote $\cO^{(\theta;R)}(x) \equiv U_\theta^{(R)}\cO(x)U_{-\theta}^{(R)}$ and $\cO^{(\theta)} \equiv \cO^{(\theta;1)}$. The insertions $\cO_a^\dagger(\infty)$ in $h_{A,a}$ and $\cO_a(0)$ in $h_{B,a}$ each sit half period around the modular cylinder from their respective cuts [Fig.~\ref{fig:radial-prep}(b)]. Transporting them onto the cuts by half-period modular flow---writing $\cO_a^\dagger(\infty) = U_\pi^{(\xi)}(\cO_a^\dagger)^{(-\pi;\xi)}(\infty)U_{-\pi}^{(\xi)}$ and $\cO_a(0) = U_\pi\cO_a^{(-\pi)}(0)U_{-\pi}$---and using $U_\theta^{(R)}\ket{0} = \ket{0}$ yields the radial sewing expansion in sandwiched form:
\begin{equation}
  \begin{aligned}
    \rho_{AB} = \sum_a
      &\bigl(\rho_A^{1/2}\, (\cO_a^\dagger)^{(-\pi;\xi)}(\infty)\, \rho_A^{1/2}\bigr) \\
      &\otimes
      \bigl(\rho_B^{1/2}\, \cO_a^{(-\pi)}(0)\, \rho_B^{1/2}\bigr).
  \end{aligned}
  \label{eq:rhoAB-exp}
\end{equation}

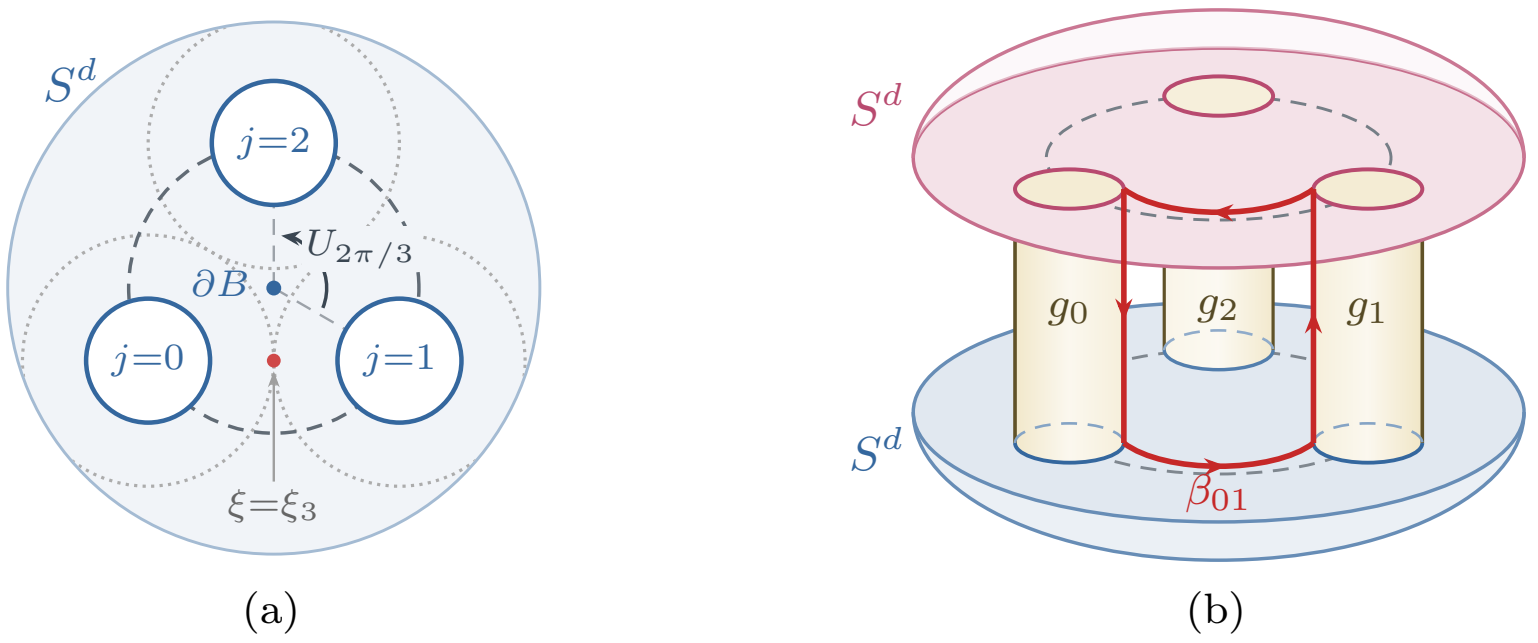
\begin{figure*}[tbp]
  \centering
  \input{figures/sewing}
  \caption{Sewn manifold geometry for $\SRMI{n}$ (shown for $n=3$). (a) The $n$ excised caps on $S^d$ lie evenly along the modular great circle (dashed) and collide at $\xi = \xi_n$ (dotted). (b) Gluing two copies by the radial cylinders $g_j$ (gold) produces $M_{n-1}(\xi)$ with thermal cycles $\beta_{ij}$ (crimson).}
  \label{fig:sewing}
\end{figure*}

The key simplification is that conjugating Eq.~\eqref{eq:rhoAB-exp} by $\sigma^{\frac{1-n}{2n}} = (\rho_A\otimes\rho_B)^{\frac{1-n}{2n}}$ gives
\begin{equation}
  \begin{aligned}
    \sigma^{\frac{1-n}{2n}}\rho_{AB}\sigma^{\frac{1-n}{2n}}
    = \sum_a
      &\bigl(\rho_A^{\frac{1}{2n}}\, (\cO_a^\dagger)^{(-\pi;\xi)}(\infty)\, \rho_A^{\frac{1}{2n}}\bigr) \\
      &\otimes
      \bigl(\rho_B^{\frac{1}{2n}}\, \cO_a^{(-\pi)}(0)\, \rho_B^{\frac{1}{2n}}\bigr),
  \end{aligned}
  \label{eq:cancellation}
\end{equation}
with each $\rho^{\frac{1}{2n}}_{A,B}$ factor implementing a modular rotation by $\pi/n$. Taking the $n$-fold cyclic trace therefore corresponds to having $n$ operator insertions equally spaced in modular angle by $2\pi/n$ [Fig.~\ref{fig:sewing}(a)]. Using $(\cO_a^\dagger)^{(\theta;\xi)}(\infty) = \xi^{\Delta_a}\xi^D(\cO_a^\dagger)^{(\theta)}(\infty)\xi^{-D}$ and vacuum invariance, we express the SRMI as a sum of unit-ball modular-flowed vacuum correlators:
\begin{equation}
  \begin{aligned}
    e^{(n-1)\SRMI{n}(A:B)}
    &= \sum_{\bm{a}} \xi^{\sum_j \Delta_{a_j}}
      \Bigl\langle\prod_{j=0}^{n-1}(\cO_{a_j}^\dagger)^{(-\theta_j)}(\infty)\Bigr\rangle \\
    &\qquad\qquad\quad\times
      \Bigl\langle\prod_{j=0}^{n-1}\cO_{a_j}^{(\theta_j)}(0)\Bigr\rangle \\
    &= \sum_{\bm{a}} \xi^{\sum_j \Delta_{a_j}}
      \Bigl|\Bigl\langle\prod_{j=0}^{n-1}\cO_{a_j}^{(\theta_j)}(0)\Bigr\rangle\Bigr|^2,
  \end{aligned}
  \label{eq:sewing-sum}
\end{equation}
with $\theta_j = 2\pi j/n$. The second equality follows from $(P_0 + K_0)^\dagger = P_0 + K_0$ in radial quantization, which implies $[\cO_a^{(\theta_j)}(0)]^\dagger = (\cO_a^\dagger)^{(-\theta_j)}(\infty)$.

Equation~\eqref{eq:sewing-sum} is the state-sum decomposition of a CFT partition function on a compact manifold $M_{n-1}(\xi) \simeq \mathop{\#}^{n-1}(S^1\times S^{d-1})$ obtained by gluing two $n$-punctured copies of $S^d = \mathbb{R}^d \cup \{\infty\}$ with $n$ cylinders [Fig.~\ref{fig:sewing}(b)], generalizing the genus-two ($n=3$) construction of Ref.~\cite{BLOSD}. In radial quantization, contracting $\cO_a^\dagger(\infty)$ on $S^d_A$ with $\xi^{\Delta_a}\cO_a(0)$ on $S^d_B$ glues the two spheres by the dilation $x_B = \xi^D\cdot x_A = \xi x_A$. For the $j$-th cylinder in Eq.~\eqref{eq:sewing-sum}, the insertions are rotated by $U_{-\theta_j}$ on $S^d_A$ and $U_{\theta_j}$ on $S^d_B$, giving the gluing map $x_B = U_{\theta_j}\xi^D U_{\theta_j}\cdot x_A$. We now apply the global radial inversion $x_A' = I(x_A) = x_A/x_A^2$ to $S^d_A$. Using $U_{\theta_j} I = I U_{-\theta_j}$ from Eq.~\eqref{eq:gtheta}, this casts all $n$ transition maps $x_B = g_j\cdot x_A'$ into the orientation-reversing component $G^-$ of $O(d+1,1)$~\cite{BLOSD}:
\begin{equation}
  g_j = U_{\theta_j}\,\xi^D I\, U_{-\theta_j} \in G^-,
  \qquad j = 0,\dots,n-1,
  \label{eq:seam-gluing}
\end{equation}
where each $g_j$ is an involution satisfying $g_j^2 = \id$.

As in Ref.~\cite{BLOSD}, the flat conformal structure of $M_{n-1}(\xi)$ is determined invariantly by $(g_0,\dots,g_{n-1}) \in G\backslash(G^-)^n/G$ modulo independent conformal transformations on the two spheres. Its gauge-invariant pairwise holonomies $g_i^{-1}g_j = g_i g_j \in SO^+(d+1,1)$ act nontrivially only in an $SO^+(2,1)$ three-plane, with zero transverse angular twists. Their eigenvalues $(e^{\pm\beta_{ij}}, 1, \dots, 1)$ are fixed by $\Tr_{SO(d+1,1)}(g_i^{-1}g_j) = 2\cosh\beta_{ij} + d$, yielding effective inverse temperatures
\begin{equation}
  \begin{aligned}
    \beta_{ij}(\xi) &= 4\arccosh\!\left[\frac{(1+\xi)\sin(\pi|i-j|/n)}{2\sqrt{\xi}}\right], \\
    \beta_n(\xi) &\equiv \beta_{01}(\xi).
  \end{aligned}
  \label{eq:beta-ij}
\end{equation}
The glued manifold remains smooth while every effective inverse temperature is real and positive, $\beta_{ij}(\xi) > 0$. Its boundary is the infinite-temperature limit of the adjacent-pair inverse temperature $\beta_n(\xi) \to 0$:
\begin{equation}
  0 < \xi < \xi_n = \tan^2\frac{\pi}{2n}
  \iff
  0 < \eta < \eta_n = \sin^2\frac{\pi}{n}.
  \label{eq:domain}
\end{equation}

For $M_{n-1}(\xi)$ endowed with a metric, the raw path-integral partition function $Z[M_{n-1}(\xi)]$ differs from the state sum Eq.~\eqref{eq:sewing-sum} by an overall metric-dependent prefactor encoding the Weyl anomaly, Casimir energy, and gluing normalizations~\cite{BLOSD}. In Eq.~\eqref{eq:sewing-sum}, the single-vacuum term $\bm{a} = (0,\dots,0)$---where only the vacuum state $\ket{0}$ propagates through every cylinder---equals unity. Physically, keeping only $a = 0$ in Eq.~\eqref{eq:rhoAB-exp} reduces $\rho_{AB}$ to the factorized state $\rho_A\otimes\rho_B$, for which $\SRMI{n} = 0$. That overall prefactor therefore coincides with the single-vacuum contribution $Z_{0,\dots,0}[M_{n-1}(\xi)]$. Setting $\Zhat[M_{n-1}(\xi)] \equiv Z[M_{n-1}(\xi)]/Z_{0,\dots,0}[M_{n-1}(\xi)]$ cancels all metric dependence and yields our central identity:
\begin{equation}
  \boxed{
    \SRMI{n}(A:B) = \frac{1}{n-1}\log\Zhat[M_{n-1}(\xi)],
    \quad 0 < \xi < \xi_n.
  }
  \label{eq:master}
\end{equation}

\emph{Order two: thermal partition functions}---At $n=2$, we have $\xi_2 = 1$ and effective inverse temperature $\beta_2(\xi) = -2\log\xi$, with manifold $M_1(\xi) \simeq S^1_{\beta_2} \times S^{d-1}$. For all $0 < \xi < 1$, Eq.~\eqref{eq:master} therefore identifies the second SRMI with the thermal partition function on $S^{d-1}$ at inverse temperature $\beta_2$. Monotonicity in the R\'enyi order immediately yields an upper bound on MI:
\begin{equation}
  \boxed{
    I(A:B) \le \SRMI{2}(A:B) = \log Z_{S^{d-1}}(\beta_2),
  }
  \label{eq:n2-thermal}
\end{equation}
where $Z_{S^{d-1}}(\beta) = \Tr_{\cH_{S^{d-1}}} e^{-\beta D}$. As $\xi \to 1$, $\beta_2(\xi) \sim 2(1-\xi) \to 0$ enters the high-temperature regime of thermal partition function $\log Z_{S^{d-1}}(\beta) \sim f\Vol(S^{d-1})\,\beta^{-(d-1)}$, giving the contact divergence
\begin{equation}
  \SRMI{2}(\xi) \sim \frac{f\Vol(S^{d-1})}{\beta_2(\xi)^{d-1}} \sim \frac{f\Vol(S^{d-1})}{[2(1-\xi)]^{d-1}},
  \label{eq:n2-contact}
\end{equation}
where $-f$ is the flat-space thermal free-energy density at unit temperature ($f = \pi c/6$ in $d = 2$). This divergence is strictly stronger than the $(1-\xi)^{-(d-2)}$ area law of $I(A:B)$, or $\log(1-\xi)^{-1}$ in $d = 2$~\cite{Casini:2005zv,Cardy:2013nua}.

The same radial construction extends directly to two relatively boosted spacelike balls, specified conformally by their proper-radius ratio $\xi$ and relative rapidity $\chi$ with $\xi e^{|\chi|} < 1$. In Euclidean signature, rotating $A$ relative to $B$ by angle $\alpha$ in the $(\tau,x^1)$ plane replaces the annulus propagator $\xi^D$ by $\xi^D e^{i\alpha J_{01}}$, where $J_{01} = -i M_{01}$ is the Hermitian $(\tau,x^1)$ rotation generator. Each gluing map in Eq.~\eqref{eq:seam-gluing} accordingly becomes $g_j = U_{\theta_j}\,\xi^D e^{i\alpha J_{01}} I\, U_{-\theta_j}$. At $n = 2$, the half-period map $U_\pi$ combines inversion with $\tau \mapsto -\tau$, flipping $J_{01} \mapsto -J_{01}$ and giving the single holonomy $g_1^{-1} g_0 = \xi^{-2D} e^{2i\alpha J_{01}}$. Continuing $\alpha = -i\chi$ therefore identifies $\SRMI{2}$ with the thermal partition function on $S^{d-1}$ with angular potential $\omega_2$:
\begin{equation}
  \boxed{
    I(A:B) \le \SRMI{2}(A:B) = \log Z_{S^{d-1}}(\beta_2,\omega_2),
  }
  \label{eq:n2-boosted}
\end{equation}
where $Z_{S^{d-1}}(\beta,\omega) = \Tr_{\cH_{S^{d-1}}} e^{-\beta(D - \omega J_{01})}$, and $\omega_2(\xi,\chi) = -\chi/\log\xi \in (-1,1)$.

Equation~\eqref{eq:n2-boosted} strictly improves a similar thermal bound on MI in Ref.~\cite{CeyhanFaulknerThermal}, which in our notation reads $I(A:B) \le \log 2 + 2\log Z_{S^{d-1}}(\beta_2/2,\omega_2)$. Our bound drops both the additive $\log 2$ and the prefactor $2$, and doubles the inverse temperature from $\beta_2/2$ to $\beta_2$ at fixed $\omega_2$.

\begin{figure}[tbp]
  \centering
  \includegraphics[width=\columnwidth]{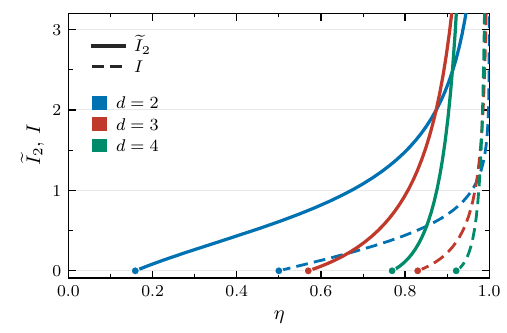}
  \caption{Holographic $\SRMI{2}(\eta)$ (solid) and $I(\eta)$ (dashed), in units of $\Vol(S^{d-1})/(4G_N)$, for $d = 2, 3, 4$. Dots mark the Hawking--Page and RT transitions ($\eta_{\mathrm{HP}} < \eta_{\mathrm{RT}}$).}
  \label{fig:holography}
\end{figure}

For a holographic \CFT{d} in the classical gravity approximation, the thermal partition function in Eq.~\eqref{eq:n2-thermal} is evaluated by the on-shell action of Euclidean bulk saddles: thermal \AdS{d+1} in the confined phase and Schwarzschild--\AdS{d+1} in the deconfined phase~\cite{HawkingPage,WittenThermal}. The bound in Eq.~\eqref{eq:n2-thermal} implies that the Hawking--Page transition at $\beta_{\mathrm{HP}} = 2\pi/(d-1)$ must precede the connected--disconnected Ryu--Takayanagi (RT) phase transition of $I(A:B)$~\cite{RyuTakayanagi,FondaGiomiSalvioTonni}, yielding the geometric bound
\begin{equation}
  \begin{aligned}
    \xi_{\mathrm{RT}} &\ge \xi_{\mathrm{HP}} = e^{-\pi/(d-1)}, \\
    \eta_{\mathrm{RT}} &\ge \eta_{\mathrm{HP}} = \sech^2\!\left(\frac{\pi}{2(d-1)}\right).
  \end{aligned}
  \label{eq:hp-exact}
\end{equation}
Figure~\ref{fig:holography} compares the holographic curves of $\SRMI{2}(\eta)$ and $I(\eta)$, confirming the monotonicity in R\'enyi order, $\SRMI{2}(\eta) \ge I(\eta)$, in addition to the transition ordering $\eta_{\mathrm{HP}} < \eta_{\mathrm{RT}}$.

\emph{Higher orders: finite-separation divergence and hot-spot limit}---Because every summand in Eq.~\eqref{eq:sewing-sum} is nonnegative, for any integer $n \ge 3$ we may project $n-2$ cylinders to the vacuum state $\ket{0}$ while summing over all states on an adjacent pair. This leaves two spheres joined by two cylinders, conformally $S^1_{\beta_n}\times S^{d-1}$, yielding the universal lower bound
\begin{equation}
  \SRMI{n}(\xi) \ge \frac{1}{n-1}\log Z_{S^{d-1}}(\beta_n).
  \label{eq:two-seam-bound}
\end{equation}
As $\xi \to \xi_n < 1$, Eq.~\eqref{eq:beta-ij} gives $\beta_n(\xi) \propto \sqrt{\xi_n - \xi} \to 0$, so $\SRMI{n}(\xi)$ diverges at finite separation, and hence for all $\xi \ge \xi_n$ by monotonicity. Since $\xi_n \to 0$ as $n \to \infty$, monotonicity in R\'enyi order immediately implies that the max-mutual information~\cite{Datta} diverges at all separations $0 < \xi < 1$:
\begin{equation}
 \SRMI{\infty}(A:B) = +\infty.
  \label{eq:dinfty}
\end{equation}

\begin{figure*}[tbp]
  \centering
  \includegraphics{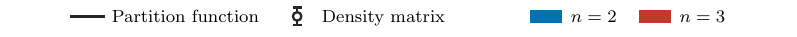}\\[-2pt]
  \subfloat[2d Ising\label{fig:comp-ising}]{\includegraphics{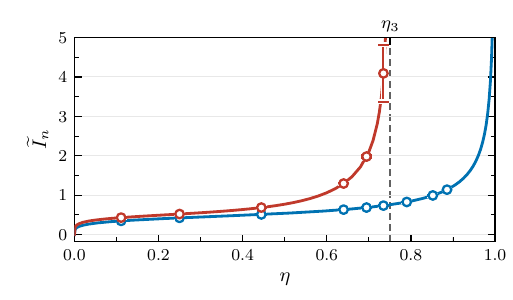}}\hfill
  \subfloat[2d compact boson ($R = \sqrt{2}$)\label{fig:comp-boson}]{\includegraphics{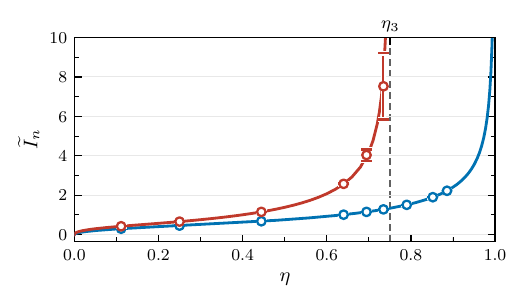}}\\[4pt]
  \subfloat[3d free scalar\label{fig:comp-scalar3}]{\includegraphics{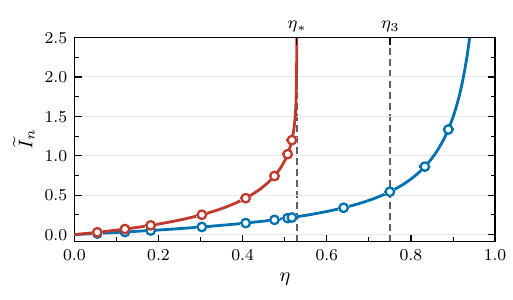}}\hfill
  \subfloat[5d free scalar\label{fig:comp-scalar5}]{\includegraphics{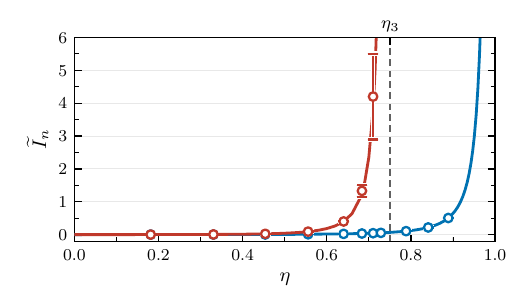}}
  \caption{Comparison between CFT partition functions (solid curves) and reduced density matrix results (open circles) for $n = 2$ and $n = 3$. Vertical dashed lines mark the finite-separation divergence of $\SRMI{3}$ at $\eta_3 = 3/4$, and at the earlier $d = 3$ Yamabe threshold $\eta_* \approx 0.5287$ in (c). Error bars indicate extrapolation uncertainties.}
  \label{fig:comparison}
\end{figure*}

For a generic interacting \CFT{d} whose compactification on $S^1_\beta$ is gapped, $\SRMI{n}(\xi)$ is expected to remain finite below $\xi_n$ and diverge as $\xi \to \xi_n$, where the $n$ adjacent inverse temperatures vanish. Applying the hot-spot conjecture of Ref.~\cite{BLOSD} to the $n$ resulting hot spots yields
\begin{equation}
  \SRMI{n}(\xi) \sim \frac{n}{n-1}\log Z_{S^{d-1}}(\beta_n) + \frac{\log R_n}{n-1} + o(1),
  \label{eq:hotspot-general}
\end{equation}
which by Eq.~\eqref{eq:n2-contact} diverges as $(\xi_n - \xi)^{-(d-1)/2}$, with $R_n$ the finite $n$-point hot-spot correlator~\cite{BLOSD}. In $d = 2$, modular invariance establishes Eq.~\eqref{eq:hotspot-general} rigorously for any compact unitary \CFT{2} with a unique vacuum~\cite{BLOSD,KOP}. By contrast, special free theories lacking a thermal mass gap can diverge strictly before $\xi_n$, as we show below for the $d = 3$ free conformal scalar.

\emph{Explicit CFT evaluations at finite separation}---We now evaluate Eq.~\eqref{eq:master} explicitly in CFT examples and compare against density matrix calculations of SRMI.

In $d = 2$, $M_{n-1}(\xi) = (S^2 \setminus \Lambda_\xi)/\Gamma_\xi$ is a genus-$g$ Riemann surface with $g = n - 1$, uniformized by the real Fuchsian Schottky group $\Gamma_\xi \subset PSL(2,\mathbb{R})$ with limit set $\Lambda_\xi \subset \mathbb{R}\cup\{\infty\}$~\cite{Martinec}. Taking cylinder $0$ as a base, $\Gamma_\xi$ is freely generated by the $g$ pairwise holonomies $\gamma_a = g_0^{-1} g_a$ ($a = 1,\dots,g$). In the canonical homology basis of Fig.~\ref{fig:sewing}(b), $A_a$ encircles cylinder $a$, while $B_a$ passes through cylinder $0$ and returns through cylinder $a$, shown in crimson for $a = 1$. The antiholomorphic reflection exchanging $S^2_A \leftrightarrow S^2_B$ fixes each $A_a$ and reverses each $B_a$, forcing the period matrix $\Omega_g(\xi)$ to be purely imaginary. Moreover, the cyclic $\mathbb{Z}_n$ symmetry of the $n$ cylinders fixes four points---two modular-flow fixed points on each sphere---with opposite monodromies on each pair. By Riemann--Hurwitz, $M_{n-1}(\xi)/\mathbb{Z}_n \cong \mathbb{CP}^1$, so $M_{n-1}(\xi)$ is conformally equivalent to the $n$-sheeted two-interval replica surface $\cR_{n,2}(x_n(\xi))$ of Refs.~\cite{CCT09,CCT11} at a unique real cross ratio $x_n(\xi) \in (0,1)$, increasing monotonically from $x_n(\xi) = 4n^2\xi + O(\xi^2)$ to $x_n(\xi_n) = 1$. In the $(A_a,B_a)$ basis, $\Omega_g(\xi)$ takes the closed form~\cite{CCT09}
\begin{equation}
  \begin{aligned}
    (\Omega_g)_{ab}
    &= \frac{2i}{n}\sum_{k=1}^{n-1}
       \frac{\kappa_{k/n}\bigl(x_n(\xi)\bigr)}{\sin(\pi k/n)} \\
    &\quad\times
       \sin\!\frac{\pi k a}{n}\,\sin\!\frac{\pi k b}{n}\,\cos\!\frac{\pi k(a-b)}{n},
  \end{aligned}
  \label{eq:cft2-period}
\end{equation}
where $\kappa_y(x) \equiv {}_2F_1(y,1-y;1;1-x)/{}_2F_1(y,1-y;1;x)$, and $x_n(\xi)$ is fixed uniquely by matching this matrix to the Schottky period matrix of $\Gamma_\xi$~\cite{Martinec}.

Writing $\Zhat_n(\xi) \equiv \Zhat[M_{n-1}(\xi)]$, the vacuum-normalized partition function of the compact boson of radius $R$ on $M_{n-1}(\xi)$ factorizes into its oscillator determinant and zero-mode lattice sum~\cite{Martinec,McIntyreTakhtajan,DVV}:
\begin{equation}
  \Zhat_n^{\mathrm{boson}}(\xi)
  = \cD_g(\xi)\,\Theta(0|\Omega_g/R^2)\,\Theta(0|R^2\Omega_g),
  \label{eq:cft2-boson}
\end{equation}
where
\begin{equation}
  \cD_g(\xi) = \prod_{[\gamma]\,\mathrm{prim}}\prod_{m=1}^{\infty}\bigl(1 - q_\gamma^m\bigr)^{-1}
  \label{eq:cft2-schottky}
\end{equation}
is the free-boson Schottky oscillator determinant over all primitive conjugacy classes $[\gamma] \subset \Gamma_\xi$ with contracting multiplier $q_\gamma \in (0,1)$, reducing to $q_\gamma = e^{-\beta_{ij}}$ for $\gamma = g_i^{-1}g_j$. Here $\Theta[\bm{\delta}](0|\Omega_g)$ denotes the genus-$g$ Siegel theta function with half-integer characteristic $\bm{\delta} \in \{0,\frac{1}{2}\}^{2g}$, abbreviated $\Theta(0|\Omega_g)$ for $\bm{\delta} = \bm{0}$. Because $M_{n-1}(\xi)$ and $\cR_{n,2}(x_n(\xi))$ are conformally equivalent, $\Zhat_n(\xi)$ only differs from the two-interval R\'enyi scaling function $\cF_n(x_n(\xi))$ of Refs.~\cite{CCT09,CCT11} by a universal Weyl anomaly factor. Fixing this factor with the self-dual compact boson at $R = 1$, for which $\cF_n = 1$~\cite{CCT09}, yields
\begin{equation}
  \Zhat_n(\xi)
  = \Bigl[\cD_g(\xi)\,\Theta(0|\Omega_g)^2\Bigr]^c\,\cF_n\bigl(x_n(\xi)\bigr).
  \label{eq:cft2-cct}
\end{equation}
In particular, for the diagonal Ising CFT~\cite{DVV,CCT11},
\begin{equation}
  \Zhat_n^{\mathrm{Ising}}(\xi)
  = \cD_g(\xi)^{1/2}\,2^{-g}\sum_{\bm{\delta}\;\mathrm{even}}\bigl|\Theta[\bm{\delta}](0|\Omega_g)\bigr|.
  \label{eq:cft2-ising}
\end{equation}

In $d \ge 3$, the pairwise holonomies act solely in the $SO^+(2,1)$ subgroup with trivial transverse twists, so the same rank-$(n-1)$ Fuchsian Schottky group $\Gamma_\xi$ uniformizes $M_{n-1}(\xi) = (S^d \setminus \Lambda_\xi)/\Gamma_\xi$. For a real free conformal scalar in odd $d \ge 3$, the conformal-Laplacian determinant reduces via the Selberg zeta function~\cite{Guillarmou} to
\begin{equation}
  \begin{aligned}
    \log\Zhat_n^{\mathrm{scalar}}(\xi)
    &= \frac{1}{2} \sum_{[\gamma]\,\mathrm{prim}} \sum_{k=1}^\infty \frac{h_d(q_\gamma^k)}{k}, \\
    h_d(q) &= \frac{q^{(d-2)/2}(1+q)}{(1-q)^{d-1}},
  \end{aligned}
  \label{eq:selberg-scalar}
\end{equation}
summed over the same primitive classes $[\gamma] \subset \Gamma_\xi$ and multipliers $q_\gamma$ as in Eq.~\eqref{eq:cft2-schottky}. While Eq.~\eqref{eq:selberg-scalar} remains finite up to $\xi_n$ for odd $d \ge 5$, in $d = 3$ the conformal Laplacian develops a Yamabe-negative zero mode strictly before contact~\cite{IzekiNayatani}, causing $\SRMI{3}$ to diverge at $\eta_* \approx 0.5287 < \eta_3 = 3/4$.

To test Eq.~\eqref{eq:master}, we compute $\SRMI{2}$ and $\SRMI{3}$ directly from reduced density matrices. We use continuum extrapolations of critical Ising and XX spin chains in $d = 2$~\cite{FagottiCalabrese}, and continuum vacuum Gaussian covariances of the free scalar on spatial balls in $d = 3, 5$~\cite{DydaKuznetsovKwasnicki,SeshadreesanLamiWilde}. As shown in Fig.~\ref{fig:comparison}, the sewn partition functions agree well with the density matrix results in all four models.

\emph{Discussion}---Although we have used formal density matrices throughout for transparency, the sandwiched R\'enyi divergence $\SRel{\alpha}$ is rigorously defined in Type $\mathrm{III}_1$ algebraic QFT via weighted noncommutative $L_p$ spaces, where data processing and monotonicity in $\alpha$ hold directly on normal states~\cite{BertaScholzTomamichel,Jencova,Lashkari}. In this setting, the split property guarantees a normal product state at positive separation, while modular nuclearity relates conformal ball inclusions to the radial operator $\xi^D$ and yields upper bounds on mutual information~\cite{BDL,PanebiancoWegener,CeyhanFaulknerThermal}. Establishing Eq.~\eqref{eq:master} directly in this operator algebraic framework remains an open problem.

\begin{acknowledgments}
\emph{Acknowledgments}---We acknowledge support from the Tsinghua Shuimu Scholar program. We acknowledge use of ChatGPT and Gemini for assistance with derivation, computation, and manuscript preparation. We take responsibility for validity of the results.
\end{acknowledgments}

\FloatBarrier
\bibliography{refs}

\end{document}

%% file: macros.tex
\usepackage{amsmath,amssymb,mathtools,bm}
\usepackage{graphicx,xcolor,tikz,booktabs}
\usepackage{microtype}
\usepackage{placeins}
\usepackage[caption=false]{subfig}
\usetikzlibrary{arrows.meta,calc,positioning,decorations.markings}
\usepackage[colorlinks=true,linkcolor=blue,citecolor=blue,urlcolor=blue]{hyperref}
\usepackage{bookmark}

\DeclareMathOperator{\Tr}{Tr}

\DeclareMathOperator{\arccosh}{arccosh}

\DeclareMathOperator{\sech}{sech}

\DeclareMathOperator{\Vol}{Vol}

\newcommand{\id}{\mathbf{1}}
\newcommand{\ket}[1]{|#1\rangle}
\newcommand{\bra}[1]{\langle#1|}

\newcommand{\cO}{\mathcal{O}}
\newcommand{\cH}{\mathcal{H}}

\newcommand{\cD}{\mathcal{D}}
\newcommand{\cF}{\mathcal{F}}
\newcommand{\cR}{\mathcal{R}}
\newcommand{\SRMI}[1]{\widetilde{I}_{#1}}
\newcommand{\SRel}[1]{\widetilde{D}_{#1}}
\newcommand{\Zhat}{\widehat{Z}}
\newcommand{\CFT}[1]{\ensuremath{\mathrm{CFT}_{#1}}}
\newcommand{\AdS}[1]{\ensuremath{\mathrm{AdS}_{#1}}}

%% file: figures/radial-preparation.tex
% Native 2-panel TikZ subfigures deriving the radial sewing expansion of \rho_{AB}.
\subfloat[\label{fig:radial-prep-a}]{%
\begin{tikzpicture}[
  x=1cm,y=1cm,>=Stealth,font=\small,
  every node/.style={inner sep=1.5pt}
]
  \definecolor{rose}{RGB}{184,74,111}
  \definecolor{cobalt}{RGB}{52,103,158}
  \definecolor{gold}{RGB}{213,182,90}
  \definecolor{slate}{RGB}{58,70,82}

  % Top row: radial cut isolating annulus \xi < |x| < 1
  \node[font=\small,anchor=east] at (1.60,1.25) {$\rho_{AB} = \displaystyle\sum_a$};

  \fill[cobalt!9]  (1.78,0.02) rectangle (6.38,2.48);
  \fill[gold!24]   (4.08,1.25) circle (1.02);
  \fill[rose!11]   (4.08,1.25) circle (0.50);
  \draw[slate!55,line width=0.48pt] (1.78,0.02) rectangle (6.38,2.48);
  \draw[slate,line width=0.65pt]    (4.08,1.25) circle (1.02);
  \draw[slate,line width=0.65pt]    (4.08,1.25) circle (0.50);

  % Wavy cuts A and B on tau = 0
  \draw[rose,line width=1.0pt,line cap=round]
    plot[domain=3.58:4.58,samples=51] (\x,{1.25+0.028*sin(4*360*(\x-3.58)/1.00)});
  \draw[cobalt,line width=1.0pt,line cap=round]
    plot[domain=1.78:3.06,samples=51] (\x,{1.25+0.028*sin(4*360*(\x-1.78)/1.28)});
  \draw[cobalt,line width=1.0pt,line cap=round]
    plot[domain=5.10:6.38,samples=51] (\x,{1.25+0.028*sin(4*360*(\x-5.10)/1.28)});

  \fill[rose]   (3.58,1.25) circle (1.2pt) (4.58,1.25) circle (1.2pt);
  \fill[cobalt] (3.06,1.25) circle (1.3pt) (5.10,1.25) circle (1.3pt);

  \node[rose,font=\footnotesize]          at (4.08,1.50) {$A$};
  \node[cobalt,font=\footnotesize]        at (5.84,1.50) {$B$};
  \node[slate,font=\scriptsize]           at (3.12,2.04) {$\ket{a}$};
  \node[slate,font=\scriptsize]           at (3.42,1.64) {$\bra{a}$};
  \node[slate,font=\scriptsize]           at (4.08,1.89) {$|x|{=}\xi$};
  \node[slate,font=\scriptsize]           at (5.24,0.34) {$|x|{=}1$};
  \node[gold!45!black,font=\scriptsize]   at (4.08,0.50) {$\xi^D$};

  % Tiny Euclidean axes
  \draw[->,slate!80,line width=0.45pt,>={Stealth[length=2.8pt,width=2.1pt]}]
    (5.58,1.92) -- (5.58,2.26) node[above,font=\scriptsize] {$\tau$};
  \draw[->,slate!80,line width=0.45pt,>={Stealth[length=2.8pt,width=2.1pt]}]
    (5.58,1.92) -- (5.92,1.92) node[right,font=\scriptsize] {$x^1$};

  % Bottom row: = \sum_a h_{A,a} \otimes h_{B,a}
  \node[font=\small,anchor=east] at (1.60,-1.18) {$= \displaystyle\sum_a$};

  % Pink box h_{A,a}
  \fill[rose!11] (1.78,-1.88) rectangle (3.86,-0.48);
  \draw[slate!55,line width=0.48pt] (1.78,-1.88) rectangle (3.86,-0.48);
  \draw[rose,line width=1.0pt,line cap=round]
    plot[domain=1.96:2.68,samples=41] (\x,{-1.18+0.028*sin(3*360*(\x-1.96)/0.72)});
  \fill[rose] (1.96,-1.18) circle (1.2pt) (2.68,-1.18) circle (1.2pt);
  \node[rose,font=\footnotesize] at (2.32,-0.92) {$A$};
  \fill[rose] (3.24,-1.18) circle (1.7pt);
  \node[font=\scriptsize,anchor=south] at (3.24,-1.04) {$\cO_a^\dagger(\infty)$};
  \node[font=\footnotesize,anchor=north] at (2.82,-1.96) {$h_{A,a}$};

  \node[font=\small] at (4.10,-1.18) {$\otimes$};

  % Cobalt box h_{B,a}
  \fill[cobalt!9] (4.34,-1.88) rectangle (6.42,-0.48);
  \draw[slate!55,line width=0.48pt] (4.34,-1.88) rectangle (6.42,-0.48);
  \draw[cobalt,line width=1.0pt,line cap=round]
    plot[domain=4.34:4.92,samples=31] (\x,{-1.18+0.028*sin(3*360*(\x-4.34)/0.58)});
  \draw[cobalt,line width=1.0pt,line cap=round]
    plot[domain=5.84:6.42,samples=31] (\x,{-1.18+0.028*sin(3*360*(\x-5.84)/0.58)});
  \fill[cobalt] (4.92,-1.18) circle (1.2pt) (5.84,-1.18) circle (1.2pt);
  \node[cobalt,font=\footnotesize] at (6.12,-0.92) {$B$};
  \fill[cobalt] (5.38,-1.18) circle (1.7pt);
  \node[font=\scriptsize,anchor=south] at (5.38,-1.04) {$\cO_a(0)$};
  \node[font=\footnotesize,anchor=north] at (5.38,-1.96) {$h_{B,a}$};

  \path (0.15,-2.32) (6.42,2.50);
\end{tikzpicture}%
}\hspace{0.9cm}%
\subfloat[\label{fig:radial-prep-b}]{%
\begin{tikzpicture}[
  x=1cm,y=1cm,>=Stealth,font=\small,
  midarrow/.style={
    postaction={decorate},
    decoration={markings,mark=at position #1 with {\arrow{Stealth[length=4.0pt,width=3.2pt]}}}
  },
  every node/.style={inner sep=1.5pt}
]
  \definecolor{rose}{RGB}{184,74,111}
  \definecolor{cobalt}{RGB}{52,103,158}
  \definecolor{slate}{RGB}{58,70,82}

  % --- Top row: h_{A,a} on the modular cylinder of A = B_\xi ---
  \begin{scope}[shift={(0,1.25)}]
    \node[font=\small,anchor=east] at (0.00,0) {$h_{A,a} =$};

    \begin{scope}[shift={(2.05,0)}]
      % Cylinder body
      \fill[rose!11] (-0.78,-0.92) rectangle (0.78,0.92);
      \fill[rose!13] (0, 0.92) ellipse (0.78 and 0.21);
      \fill[rose!11] (0,-0.92) ellipse (0.78 and 0.21);
      \draw[slate!75,line width=0.52pt] (0.78,-0.92) -- (0.78,0.92);
      \draw[slate!75,line width=0.52pt] (0, 0.92) ellipse (0.78 and 0.21);
      \draw[slate!75,line width=0.52pt] (-0.78,-0.92)
        arc[start angle=180,end angle=360,x radius=0.78,y radius=0.21];
      \draw[slate!40,densely dashed,line width=0.45pt] (-0.78,-0.92)
        arc[start angle=180,end angle=0,x radius=0.78,y radius=0.21];

      % Unsewn cut along A at \varphi = 0, 2\pi
      \draw[rose,line width=1.0pt,line cap=round]
        plot[domain=-0.92:0.92,samples=61]
        ({-0.78+0.015*sin(4*360*(\x+0.92)/1.84)},\x);
      \node[rose,font=\scriptsize,anchor=east] at (-0.86,0.98) {$\varphi{=}0,2\pi$};

      % Positive A modular evolution: front 0 -> \pi, back \pi -> 2\pi
      \draw[slate,line width=0.62pt,midarrow=0.53] (-0.78,0)
        .. controls (-0.48,-0.36) and (0.48,-0.36) .. (0.78,0);
      \draw[slate,line width=0.62pt,densely dashed,midarrow=0.53] (0.78,0)
        .. controls (0.48,0.36) and (-0.48,0.36) .. (-0.78,0);
      \node[font=\scriptsize] at (0,-0.52) {$\rho_A^{1/2}$};
      \node[font=\scriptsize] at (0, 0.52) {$\rho_A^{1/2}$};

      % Insertion at \varphi = \pi
      \fill[rose] (0.78,0) circle (1.8pt);
      \node[font=\scriptsize,anchor=west] at (0.88, 0.16) {$\cO_a^\dagger(\infty)$};
      \node[slate,font=\scriptsize,anchor=west] at (0.88,-0.22) {$\varphi{=}\pi$};
    \end{scope}

    \node[font=\small,anchor=west] at (4.00,0)
      {$=\;\rho_A^{1/2}\,(\cO_a^\dagger)^{(-\pi;\xi)}(\infty)\,\rho_A^{1/2}$};
  \end{scope}

  % --- Bottom row: h_{B,a} on the modular cylinder of B = B_1^c ---
  \begin{scope}[shift={(0,-1.12)}]
    \node[font=\small,anchor=east] at (0.00,0) {$h_{B,a} =$};

    \begin{scope}[shift={(2.05,0)}]
      % Cylinder body
      \fill[cobalt!9]  (-0.78,-0.92) rectangle (0.78,0.92);
      \fill[cobalt!12] (0, 0.92) ellipse (0.78 and 0.21);
      \fill[cobalt!9]  (0,-0.92) ellipse (0.78 and 0.21);
      \draw[slate!75,line width=0.52pt] (-0.78,-0.92) -- (-0.78,0.92);
      \draw[slate!75,line width=0.52pt] (0, 0.92) ellipse (0.78 and 0.21);
      \draw[slate!75,line width=0.52pt] (-0.78,-0.92)
        arc[start angle=180,end angle=360,x radius=0.78,y radius=0.21];
      \draw[slate!40,densely dashed,line width=0.45pt] (-0.78,-0.92)
        arc[start angle=180,end angle=0,x radius=0.78,y radius=0.21];

      % Unsewn cut along B at \varphi = \pi, -\pi
      \draw[cobalt,line width=1.0pt,line cap=round]
        plot[domain=-0.92:0.92,samples=61]
        ({0.78+0.015*sin(4*360*(\x+0.92)/1.84)},\x);
      \node[cobalt,font=\scriptsize,anchor=west] at (0.86,0.98) {$\varphi{=}\pi,-\pi$};

      % Positive B modular evolution (decreasing \varphi): front \pi -> 0, back 0 -> -\pi
      \draw[slate,line width=0.62pt,midarrow=0.53] (0.78,0)
        .. controls (0.48,-0.36) and (-0.48,-0.36) .. (-0.78,0);
      \draw[slate,line width=0.62pt,densely dashed,midarrow=0.53] (-0.78,0)
        .. controls (-0.48,0.36) and (0.48,0.36) .. (0.78,0);
      \node[font=\scriptsize] at (0,-0.52) {$\rho_B^{1/2}$};
      \node[font=\scriptsize] at (0, 0.52) {$\rho_B^{1/2}$};

      % Insertion at \varphi = 0
      \fill[cobalt] (-0.78,0) circle (1.8pt);
      \node[font=\scriptsize,anchor=east] at (-0.88, 0.16) {$\cO_a(0)$};
      \node[slate,font=\scriptsize,anchor=east] at (-0.88,-0.22) {$\varphi{=}0$};
    \end{scope}

    \node[font=\small,anchor=west] at (4.00,0)
      {$=\;\rho_B^{1/2}\,\cO_a^{(-\pi)}(0)\,\rho_B^{1/2}$};
  \end{scope}

  \path (-1.00,-2.32) (7.50,2.50);
\end{tikzpicture}%
}

%% file: figures/sewing.tex
% Native 2-panel TikZ subfigures for integer-order SRMI sewing geometry.
\subfloat[\label{fig:sewing-a}]{%
\begin{tikzpicture}[
  x=1cm,y=1cm,>=Stealth,font=\small,
  every node/.style={inner sep=1.5pt}
]
  \definecolor{cobalt}{RGB}{52,103,158}
  \definecolor{crimson}{RGB}{198,40,40}
  \definecolor{slate}{RGB}{58,70,82}

  % Round S^d representation (soft disk)
  \fill[cobalt!7] (0,0) circle (1.76);
  \draw[cobalt!45,line width=0.55pt] (0,0) circle (1.76);
  \node[cobalt,font=\small] at (-1.34,1.38) {$S^d$};

  % Clip contact circles gently inside S^d disk
  \begin{scope}
    \clip (0,0) circle (1.75);
    % Contact threshold circles at xi = xi_3 (radius = R*sin(60 deg) = 0.831)
    \foreach \ang in {90,210,330}{
      \draw[gray!65,densely dotted,line width=0.6pt]
        (\ang:0.96) circle (0.831);
    }
  \end{scope}

  % Modular-rotation great circle (radius R = 0.96)
  \draw[slate!80,densely dashed,line width=0.65pt] (0,0) circle (0.96);

  % Radial rays to j=1 (330 deg) and j=2 (90 deg) showing 2pi/3 spacing
  \draw[slate!50,densely dashed,line width=0.45pt] (0,0) -- (90:0.96);
  \draw[slate!50,densely dashed,line width=0.45pt] (0,0) -- (330:0.96);

  % Counterclockwise modular-rotation arrow for step 2pi/3
  \draw[slate,line width=0.65pt,->,>={Stealth[length=3.6pt,width=2.8pt]}]
    (338:0.35) arc[start angle=-22,end angle=82,radius=0.35];
  \node[slate,font=\scriptsize,fill=cobalt!7,inner sep=0.6pt] at (24:0.63) {$U_{2\pi/3}$};

  % Fixed entangling sphere axis at center
  \fill[cobalt] (0,0) circle (1.4pt);
  \node[cobalt,font=\scriptsize,anchor=east] at (-0.10,0.03) {$\partial B$};

  % Contact kissing point marker on front (0,1) bridge
  \fill[crimson!85] (0,-0.48) circle (1.3pt);
  \draw[->,gray!75,line width=0.45pt,>={Stealth[length=3pt,width=2.2pt]}]
    (0,-1.28) -- (0,-0.55);
  \node[gray!80!black,font=\scriptsize,fill=cobalt!7,inner sep=1.2pt]
    at (0,-1.44) {$\xi{=}\xi_3$};

  % Three excised spherical caps at 210 deg (j=0), 330 deg (j=1), 90 deg (j=2)
  \foreach \ang/\lbl in {210/{0},330/{1},90/{2}}{
    \fill[white] (\ang:0.96) circle (0.41);
    \draw[cobalt,line width=0.85pt] (\ang:0.96) circle (0.41);
    \node[cobalt,font=\scriptsize] at (\ang:0.96) {$j{=}\lbl$};
  }

  \path (-1.80,-1.84) (1.80,1.86);
\end{tikzpicture}%
}\hspace{1.9cm}%
\subfloat[\label{fig:sewing-b}]{%
\begin{tikzpicture}[
  x=1cm,y=1cm,>=Stealth,font=\small,
  midarrow/.style={
    postaction={decorate},
    decoration={markings,mark=at position #1 with {\arrow{Stealth[length=4.2pt,width=3.2pt]}}}
  },
  every node/.style={inner sep=1.5pt}
]
  \definecolor{rose}{RGB}{184,74,111}
  \definecolor{cobalt}{RGB}{52,103,158}
  \definecolor{gold}{RGB}{213,182,90}
  \definecolor{crimson}{RGB}{198,40,40}
  \definecolor{slate}{RGB}{58,70,82}

  % Geometry constants (foreshortening e = 0.36):
  % Top sheet z_top = +0.86, bottom sheet z_bot = -0.82
  % Outer disk: Rx = 2.02, Ry = 0.727; Dome/bowl outer Ry = 0.98
  % Modular great circle: Rg = 1.14, Rgy = 0.410
  % Cylinders (gold-tinted to match annulus \xi^D): rc = 0.36, rcy = 0.130

  % 1. Bottom sphere S^d_B (semi-elliptical bowl below + gluing disk)
  \fill[cobalt!10]
    (-2.02,-0.82) arc[start angle=180,end angle=360,x radius=2.02,y radius=0.98]
    arc[start angle=0,end angle=180,x radius=2.02,y radius=0.727] -- cycle;
  \fill[cobalt!15] (0,-0.82) ellipse (2.02 and 0.727);
  \draw[cobalt!75,line width=0.65pt]
    (-2.02,-0.82) arc[start angle=180,end angle=360,x radius=2.02,y radius=0.98];
  \draw[cobalt!75,line width=0.65pt]
    (0,-0.82) ellipse (2.02 and 0.727);

  % Bottom modular great circle on S^d_B
  \draw[slate!65,densely dashed,line width=0.55pt]
    (0,-0.82) ellipse (1.14 and 0.410);

  % 2. Back cylinder g_2 at (0, +0.410) (softer gold shading for depth)
  \shade[left color=gold!14,right color=gold!30,middle color=gold!8]
    (-0.36,-0.410) -- (-0.36,1.270)
    arc[start angle=180,end angle=360,x radius=0.36,y radius=0.130]
    -- (0.36,-0.410)
    arc[start angle=0,end angle=-180,x radius=0.36,y radius=0.130] -- cycle;
  \draw[gold!45!black,line width=0.6pt] (-0.36,-0.410) -- (-0.36,1.270);
  \draw[gold!45!black,line width=0.6pt] ( 0.36,-0.410) -- ( 0.36,1.270);
  \draw[cobalt!85,line width=0.65pt]
    (-0.36,-0.410) arc[start angle=180,end angle=360,x radius=0.36,y radius=0.130];
  \draw[cobalt!55,densely dashed,line width=0.5pt]
    (-0.36,-0.410) arc[start angle=180,end angle=0,x radius=0.36,y radius=0.130];

  % 3. Front-left cylinder g_0 at (-0.987, -0.205)
  \shade[left color=gold!32,right color=gold!22,middle color=gold!10]
    (-1.347,-1.025) -- (-1.347,0.655)
    arc[start angle=180,end angle=360,x radius=0.36,y radius=0.130]
    -- (-0.627,-1.025)
    arc[start angle=0,end angle=-180,x radius=0.36,y radius=0.130] -- cycle;
  \draw[gold!45!black,line width=0.68pt] (-1.347,-1.025) -- (-1.347,0.655);
  \draw[gold!45!black,line width=0.68pt] (-0.627,-1.025) -- (-0.627,0.655);
  \draw[cobalt,line width=0.72pt]
    (-1.347,-1.025) arc[start angle=180,end angle=360,x radius=0.36,y radius=0.130];
  \draw[cobalt!60,densely dashed,line width=0.5pt]
    (-1.347,-1.025) arc[start angle=180,end angle=0,x radius=0.36,y radius=0.130];

  % 4. Front-right cylinder g_1 at (+0.987, -0.205)
  \shade[left color=gold!22,right color=gold!34,middle color=gold!10]
    (0.627,-1.025) -- (0.627,0.655)
    arc[start angle=180,end angle=360,x radius=0.36,y radius=0.130]
    -- (1.347,-1.025)
    arc[start angle=0,end angle=-180,x radius=0.36,y radius=0.130] -- cycle;
  \draw[gold!45!black,line width=0.68pt] (0.627,-1.025) -- (0.627,0.655);
  \draw[gold!45!black,line width=0.68pt] (1.347,-1.025) -- (1.347,0.655);
  \draw[cobalt,line width=0.72pt]
    (0.627,-1.025) arc[start angle=180,end angle=360,x radius=0.36,y radius=0.130];
  \draw[cobalt!60,densely dashed,line width=0.5pt]
    (0.627,-1.025) arc[start angle=180,end angle=0,x radius=0.36,y radius=0.130];

  % 5. Top sphere S^d_A (gluing sheet + dashed great circle + 3 portals + dome)
  \fill[rose!16] (0,0.86) ellipse (2.02 and 0.727);
  \draw[slate!70,densely dashed,line width=0.55pt]
    (0,0.86) ellipse (1.14 and 0.410);
  % Three gold-shaded portals opening downward into cylinders
  \fill[gold!22] (0,1.270) ellipse (0.36 and 0.130);
  \fill[gold!26] (-0.987,0.655) ellipse (0.36 and 0.130);
  \fill[gold!26] ( 0.987,0.655) ellipse (0.36 and 0.130);

  % Crisp rose rims on top sheet & three portals
  \draw[rose!80,line width=0.65pt] (0,0.86) ellipse (2.02 and 0.727);
  \draw[rose,line width=0.75pt] (0,1.270) ellipse (0.36 and 0.130);
  \draw[rose,line width=0.75pt] (-0.987,0.655) ellipse (0.36 and 0.130);
  \draw[rose,line width=0.75pt] ( 0.987,0.655) ellipse (0.36 and 0.130);

  % Subtle translucent upper dome silhouette for S^d_A (true semi-ellipse)
  \fill[rose!6,fill opacity=0.55]
    (-2.02,0.86) arc[start angle=180,end angle=0,x radius=2.02,y radius=0.98]
    arc[start angle=0,end angle=180,x radius=2.02,y radius=0.727] -- cycle;
  \draw[rose!75,line width=0.65pt]
    (-2.02,0.86) arc[start angle=180,end angle=0,x radius=2.02,y radius=0.98];

  % 6. Crimson pinching thermal circle \beta_{01}(\xi) (from g_0^{-1} g_1)
  \draw[crimson,line width=1.05pt,midarrow=0.52]
    (-0.627,0.655) -- (-0.627,-1.025);
  \draw[crimson,line width=1.05pt,midarrow=0.54]
    (-0.627,-1.025) .. controls (-0.36,-1.23) and (0.36,-1.23) .. (0.627,-1.025);
  \draw[crimson,line width=1.05pt,midarrow=0.52]
    (0.627,-1.025) -- (0.627,0.655);
  \draw[crimson,line width=1.05pt,midarrow=0.54]
    (0.627,0.655) .. controls (0.36,0.45) and (-0.36,0.45) .. (-0.627,0.655);

  % 7. Labels
  \node[rose,font=\small]                at (-2.26,1.22) {$S^d$};
  \node[cobalt,font=\small]              at (-2.26,-1.06) {$S^d$};
  \node[gold!40!black,font=\footnotesize] at (-0.987,-0.16) {$g_0$};
  \node[gold!40!black,font=\footnotesize] at ( 0.987,-0.16) {$g_1$};
  \node[gold!40!black,font=\footnotesize] at ( 0.000,-0.14) {$g_2$};
  \node[crimson,font=\footnotesize]      at ( 0.000,-1.36) {$\beta_{01}$};

  \path (-2.55,-1.84) (2.55,1.86);
\end{tikzpicture}%
}